\documentclass[twocolumn]{aastex631}

\graphicspath{figures/}
\usepackage{amsmath}

\newcommand{\mbh}{M_{\rm BH}}
\newcommand{\led}{\lambda_{\rm Edd}}

\newcommand{\R}{\mathcal{R}}

\begin{document}

\title{The redshift evolution of the $M_{\rm BH}-M_*$ scaling relation: new insights from cosmological simulations and semi-analytic models}

\author[0000-0002-7941-1149]{Shashank Dattathri}
\affiliation{Department of Astronomy, Yale University, 219 Prospect Street, New Haven, CT 06511, USA}

\author[0000-0002-5554-8896]{Priyamvada Natarajan}
\affiliation{Department of Astronomy, Yale University, 219 Prospect Street, New Haven, CT 06511, USA}
\affiliation{Department of Physics, Yale University, 217 Prospect Street, New Haven, CT 06511, USA}
\affiliation{Black Hole Initiative at Harvard University, 20 Garden Street, Cambridge, MA 02138, USA}

\author[0000-0002-1996-0445]{Antonio J. Porras-Valverde}
\affiliation{Department of Astronomy, Yale University, 219 Prospect Street, New Haven, CT 06511, USA}

\author[0000-0001-9947-6911]{Colin J. Burke}
\affiliation{Department of Astronomy, Yale University, 219 Prospect Street, New Haven, CT 06511, USA}

\author{Nianyi Chen}
\affiliation{Department of Physics, McWilliams Center for Cosmology, Carnegie Mellon University, Pittsburgh, PA 15213, USA}

\author{Tiziana Di Matteo}
\affiliation{Department of Physics, McWilliams Center for Cosmology, Carnegie Mellon University, Pittsburgh, PA 15213, USA}
\affiliation{NSF AI Planning Institute for Physics of the Future, Carnegie Mellon University, Pittsburgh, PA 15213, USA}

\author{Yueying  Ni}
\affiliation{Center for Astrophysics–Harvard and Smithsonian, 60 Garden Street, Cambridge, MA 02138, USA}

\begin{abstract}
We study the co-evolution of black holes (BHs) and their host galaxies in the {\sc Astrid} and {\sc Illustris-TNG300} cosmological simulations and the {\sc Dark Sage} Semi-Analytic Model (SAM), focusing on the evolution of the BH mass - stellar mass ($\mbh-M_*$) relation. Due to differences in the adopted sub-grid modeling of BH seeding, dynamics, and feedback, the models differ in their predicted redshift evolution of the $\mbh-M_*$ relation. We find that it is the interplay between the star formation rate (SFR) and the black hole accretion rate (BHAR) which drives the evolution of the mean relation. We define a quantity $\mathcal{R}$, the ratio between the specific BHAR and SFR (i.e. $\mathcal{R} \equiv\ $sBHAR/sSFR), and demonstrate that it is $\mathcal{R}$ that governs the evolution of individual sources in the $\mbh-M_*$ plane. The efficiency of BH growth versus stellar mass growth in the sSFR-sBHAR plane reflects the partitioning of gas between fueling star formation versus BH accretion. This partitioning depends on the implementation of BH dynamics and the nature of how AGN feedback quenches galaxies. In the cosmological simulations ({\sc Astrid} and {\sc Illustris-TNG300}), the BHAR and SFR are intrinsically linked, resulting in a tight $\mbh-M_*$ correlation, while the {\sc Dark Sage} SAM produces a significantly larger scatter. We discuss these results in the context of recently discovered over-massive BHs and massive quenched galaxies at high redshift by the James Webb Space Telescope. 
\end{abstract}


\section{Introduction} \label{sec:intro}

Most if not all massive galaxies appear to harbor a supermassive black hole (SMBH) in their nucleus \citep{Kormendy1995}. The masses of these central black holes (BHs) correlate with several key properties of their host bulges: masses, velocity dispersion and luminosities \citep{Magorrian1998, Ferrarese2000, Tremaine2002, Kormendy2013}. The origin and evolution of these scaling relations is poorly understood at present, and the debate continues about whether they stem from ``nature" (initial conditions) or ``nurture" (subsequent growth over cosmic time). Since both SMBH feeding and star formation rely on gas reservoirs in galactic nuclei, it not surprising that these correlations exist. In fact, from the earliest attempts to incorporate BH growth into the standard framework of structure formation \citep{Haehnelt1998}, self-regulation of BH feeding and feedback has been argued as leading naturally to co-evolution of the BH and the stellar component in galactic nuclei. Subsequently, observational support for the idea of co-evolution has grown. Yet the origin of this connection, and whether it is a consequence of early BH seeding or later self-regulated growth, remains elusive. Heavy seeding models like direct collapse of pristine gas in the very early halos \citep{Lodato2006,Lodato2007} predict a natural connection between the potential well depth and seed mass, and hence this formation channel offers an ab-initio explanation. With the recent discovery of UHZ1 at $z=10.1$ by the James Webb Space Telescope and the Chandra Space Telescope, there is new evidence for this channel operating in the early Universe \citep{Bogdan2024,Natarajan2024}. However, such a correlation would only be set up for heavy seeds that are expected to form more rarely than the more ubiquitous light seeds that are expected to form from the remnants of the first generation of stars. 
\par
The key to understanding these empirically determined scaling relations lies in the connection between the gas supply to the BH horizon and the gas that powers star formation. Modelling the gas flows between these vast range of scales in a galactic nucleus is a significant challenge. Recent computational breakthroughs in general relativistic magnetohydrodynamic (GRMHD) simulations (see for instance, \citep{Cho2023,Cho2024}) that successfully connect sub-pc scale accretion onto the BH with kpc scales suggest that the coupling of BH feeding and feedback that modulates star formation, might in fact be the causal explanation for this connection.
\par
Despite significant progress on calibrating the black hole-galaxy scaling relations \citep{Kormendy2013, McConnell2013, Reines2015, Greene2020}, the normalization and scatter in observed $\mbh-M_*$ relations depends on the sample definition. For example, low redshift early-type galaxies have systematically larger $\mbh/M_*$ ratios than late-type galaxies and AGNs \citep{Reines2015,Sahu2019,Greene2020}. These uncertainties lead to ambiguity in whether the $\mbh-M_*$ relation is roughly constant over redshift (as argued by e.g. \citealt{Suh2020,Li2023}) or if it evolves downward with time (as argued by e.g. \citealt{Merloni2010, Bennert2011, Zhang2023}). At high redshifts, whether observed over-massive BHs \citep{Pacucci2023,Mezcua2024,Burke2024} reflect an intrinsic evolution of the mean $\mbh-M_*$ relation or a selection bias remains controversial \citep{Lauer2007,Shen2010,Li2024}. 
\par
Nevertheless, various models attempt to link changes in gas availability or mean cosmic SFR density with redshift to an evolving $\mbh-M_*$ relation \citep{Wyithe2003, Caplar2018, Yang2018, Pacucci2024}. Given the common gas reservoir powering both star formation and black hole growth and the correlation between SFR and stellar mass in regular galaxies (as noted in the ``star-forming main sequence''), it is imperative to measure correlations between black hole accretion rate and star formation rate (e.g. \citealt{Lutz2008,Shao2010, Bonfield2011, Rosario2012, Harrison2012, Mullaney2012, Yang2017, Carraro2020}) or other properties that correlate with SFR (e.g., \citealt{Beifiori2012,Mutlu-Pakdil2018,Dullo2020}). However, the vastly different characteristic timescales of star formation and black hole accretion, in addition to the difficulties in measuring star formation rates in AGN, gives rise to a large scatter in the observed trends \citep{Hickox2014}.
\par
In this paper, we quantify the $\mbh-M_*$ scaling relation and its redshift evolution in  two independent cosmological simulations, {\sc Astrid} and {\sc IllustrisTNG}, and in the {\sc Dark Sage} semi-analytic model. Since these models use different seeding prescriptions, feedback treatments, and BH dynamics, it is useful to compare their BH-galaxy co-evolution predictions. Although the simulations only trace the fate of the gas at the Bondi radius and beyond, they contain sub-grid recipes for both BH feeding, feedback, and star formation that permits us to track the mass assembly of the BH and stars for a large population of galaxies over cosmic time. {\sc Dark Sage} on the other hand deploys multiple growth channels for BHs and implements BH feeding, feedback and star formation in a qualitatively different manner \citep{PorrasValverde2023}, yielding a wide range of BH mass assembly histories (Porras-Valverde et al. submitted).
\par
Our paper is organized as follows. In section \ref{sec:datasets}, we describe the key features of {\sc Astrid}, {\sc IllustrisTNG}, and {\sc Dark Sage}. In section \ref{sec:critera} we outline our selection criteria to make meaningful comparisons across the models. In section \ref{sec:mbh_mstar} we present the key results regarding the redshift evolution of the $\mbh-M_*$ relation in the different models. We then trace this co-evolution to the relative rates of BH accretion vs star formation in section \ref{sec:bhar_sfr}. The implications of these results are discussed in section \ref{sec:discussion} and conclusions are presented in section \ref{sec:conclusions}.

\section{Simulation and semi-analytic model Datasets} \label{sec:datasets}

\subsection{The IllustrisTNG300-1 simulation}\label{Methodology_TNG}

{\sc IllustrisTNG} is a cosmological hydrodynamical simulation suite that accounts for the astrophysical processes of star formation, gas cooling, stellar evolution, stellar feedback, black hole growth and feedback \citep{Pillepich2018, Springel2018, Naiman2018, Nelson2018, Marinacci2018, Nelson2019} within the $\Lambda$ cold dark matter ($\Lambda$CDM) cosmological framework using the {\sc AREPO} code \citep{Springel2010}, which employs a magnetohydrodynamic moving-mesh approach. In this study we use the {\sc IllustrisTNG300-1} simulation (which we refer to as TNG300), as its large volume allows us to make meaningful comparisons with the other datasets.  {\sc TNG300} has a box length 302 cMpc and a dark matter particle mass resolution of $5.9 \times 10^7 M_\odot$ and initial gas particle mass of $1.1 \times 10^7 M_\odot$. Below, we summarize the implementation of the relevant physics for BH evolution and star formation. 
\par
In {\sc TNG300}, BHs are seeded with fixed seed mass $M_{\rm seed}=1.18\times 10^6 M_\odot$ in halos above a mass threshold $M_h > 7.38 \times 10^{10} M_\odot$. The BH particle is instantaneously repositioned at every time-step to the potential minimum of the galaxy, hence there are no off-center or ``wandering'' BHs by construction. BHs grow by mergers with other BHs and by accretion of gas, which is modelled using the Bondi–Hoyle–Lyttleton prescription. Any gas particle that crosses the Bondi radius, defined as $R_{\rm B} = 2\,G\,M_{\rm BH}/c_s^2$, where $c_s$ is the sound speed of the hot gas, is assumed to be accreted by the BH.
\par
{\sc TNG300} implements two modes of AGN feedback: a high accretion thermal mode (where thermal energy is injected isotropically into the surrounding gas) and a low accretion kinetic mode (where momentum is injected into the surrounding gas particles in a random direction). The transition between the two modes is given by the Eddington ratio threshold:
\begin{equation}
    \led = \rm{min} \left( 2\times 10^{-3} \times \left( \frac{\mbh}{10^8 M_\odot} \right)^2, 0.1 \right)
\end{equation}
The kinetic feedback mode is primarily responsible for quenching massive galaxies in TNG300 \citep{Terrazas2020}.
\par
A sub-grid model for star formation within the multi-phase interstellar medium is implemented according to the \citet{Springel2003} model. Gas above a threshold density is converted to stars following the Kennicutt-Schmidt relation. Sub-grid recicpes for Type Ia and Type II supernovae provide stellar feedback and metal return into the ISM. 
\par
For an extensive overview with details of the {\sc IllustrisTNG} model, we refer the reader to \citet{Weinberger2017,Pillepich2018}.

\subsection{The {\sc Astrid} simulation}

{\sc Astrid} is a cosmological hydrodynamic simulation built on the the {\sc MP-Gadget} smoothed-particle hydrodynamics code. {\sc Astrid} has a box length 369 cMpc, which is comparable to TNG300. However, with a dark matter particle mass of $9.63 \times 10^6 M_\odot$, {\sc Astrid}'s resolution is $\sim 6$ times higher than TNG300. Currently {\sc Astrid} has been run down to $z=0.5$. 
\par
{\sc Astrid} implements sub-grid models for star formation, cooling, stellar feedback, metal enrichment, etc with prescriptions similar to TNG300 but with slight modifications. Notably, {\sc Astrid} includes models for inhomogenous hydrogen reionization and helium reionization with the goal of modelling high-redshift galaxy formation. The BH accretion and feedback prescriptions are implemented in a similar fashion to TNG300, but with different parameters (kinetic feedback is enabled only for $\led<0.05$, $\mbh>10^{8.5} M_\odot$, and $z<2.3$). Compared to other cosmological simulations, {\sc Astrid} improves BH modelling in two major ways. Firstly, instead of instantaneously repositioning the BH particle at the center of the potential well, {\sc Astrid} uses a sub-grid prescription for modelling dynamical friction on a massive perturber akin to what was implemented in the Romulus-suite of simulations \citep{Tremmel2015}. Following \citet{Chen2022} (see also \citealt{Tremmel2015}), the dynamical friction force $\vec{F}_{\rm DF}$ on a black hole with velocity $\vec{v}_{\rm BH}$ depends on the local stellar density by the Chandrasekhar formula:
\begin{equation}
    \vec{F}_{\rm DF} = -16 \pi^2 G^2 \mbh^2 m_a \log(\Lambda) \frac{\vec{v}_{\rm BH}}{v_{\rm BH}^3}\int_0^{v_{\rm BH}}dv_a v_a^2 f(v_a)
\end{equation}
where $m_a$ and $v_a$ are the masses and velocities of the surrounding star particles, $f(v)$ is the distribution function of stellar velocities, and $\log(\Lambda)$ is the Coulomb logarithm. This allows BHs to gradually sink towards the center of the galaxy. In addition, the BHs now have well-defined positions and velocities with respect to their host galaxies, allowing more robust predictions for merger rates and gravitational wave detections \citep{Chen2022_2}.
\par
Secondly, black hole seeding is implemented in {\sc Astrid} with a halo mass and stellar masses thresholds ($M_{h} \geq 7.38 \times 10^9 M_\odot$ and $M_* \geq 2.95 \times 10^6 M_\odot$). However, instead of a fixed seed mass, the seed masses are drawn probabilistically from a power-law distribution specified by:
\begin{equation}
    P(M_{\rm seed}) = 
    \begin{cases}
        0 \quad &M_{\rm seed} < M_{\rm min} \\
        \mathcal{N} M_{\rm seed}^{-n}  &M_{\rm min} \leq M_{\rm seed} \leq M_{\rm max} \\
        0 &M_{\rm seed} > M_{\rm max}
    \end{cases}
\end{equation}
where $M_{\rm min}=4.43 \times 10^{4} M_\odot$, $M_{\rm max}=4.43 \times 10^5 M_\odot$, $\mathcal{N}$ is a normalization constant, and $n=1$.
\par
Gas is allowed to cool via radiative and metal-line cooling \citep{Katz1996}. Cold dense gas is converted into stars using the prescription from \citet{Springel2003}, similar to {\sc TNG300}. The metallicity of the gas and star particles is tracked over time. Stellar feedback from Type II supernovae is incorporated following \citet{Okamoto2010}, where the wind speed is proportional to the one-dimensional velocity dispersion of the dark matter. 
\par
For more details about {\sc Astrid}, we refer the reader to \citet{Bird2022, Ni2022, Chen2023, Ni2024}.

\subsection{\textit{Semi-analytic} Model: {\sc {\sc Dark Sage}}}\label{Methodology_DS}

{\sc {\sc Dark Sage}} is a semi-analytic model of galaxy formation that tracks star formation and BH growth over cosmic time. The gas context of galaxies is evolved using a one-dimensional disk structure through a set of 30 equally-spaced logarithmic bins of fixed specific angular momentum \citep{Stevens+2016}. Each galaxy is assigned a stellar and cold gas disk with a magnitude and orientation of specific angular momentum that is evolved throughout the galaxy's history. These disks are used to calculate processes such as star formation rates, gas cooling, Toomre instabilities \citep{Tommre1964}, black hole growth, stellar and AGN feedback on an annulus-by-annulus basis. {\sc Dark Sage} constructs galaxies onto the merger trees retrieved from the Millennium N-body simulation \citep{Springel2005}. These merger trees are constructed using the {\sc l-halotree} \citep{Springel+2005.364}, and found using halo finders {\sc friends-of-friends} ({\sc FoF}) and {\sc subfind} algorithm \citep{Springel+2001}. {\sc {\sc Dark Sage}} uses the 684.9 Mpc box size with particle mass resolution of $1.1 \times 10^9 \mathrm{M}_{\odot}$. Millennium uses cosmological parameters from the Wilkinson Microwave Anisotropy Probe data \citep{Spergel+2003}, where $\Omega_M = 0.25$, $\Omega_{\Lambda} = 0.75$, $\Omega_b = 0.045$, $\sigma_8 = 0.9$, and $h = 0.73$. In this paper, we use the 2018 version of {\sc {\sc Dark Sage}} \citep{Stevens+2018}, which updates prescriptions on the cooling scale radius and dispersion support in disks. 
\par
For every well-resolved halo, {\sc {\sc Dark Sage}} assigns a black hole seed. Once seeded, BHs grow initially through Toomre unstable cold gas accretion, galaxy merge-driven accretion, BH-BH mergers, followed by Bondi accretion. BH growth is closely linked to the evolution of its host galaxy. BHs can grow through accretion of Toomre unstable gas triggered by secular instabilities or merger events. When galaxies undergo mergers, {\sc Dark Sage} drives 3\% of the resulting post-merger gas disk directly into the central BH (merge-driven accretion). The rest of the cold gas is added into the primary's disk. Subsequently, the disk undergoes starburst events which would dictate how much leftover unstable gas would feed the BH. In major galaxy mergers, the central black holes of the merging galaxies combine in the subsequent snapshot.
\par
{\sc Dark Sage} implements two types of feedback mechanisms: quasar mode and radio mode, each associated with a distinct type of black hole growth. Quasar mode feedback connects cold gas accretion to the interstellar medium (ISM), while radio mode feedback links hot gas accretion to the circum-galactic medium (CGM). Both feedback mechanisms are proportional to the black hole mass. Quasar mode feedback heats up the cold gas disk starting with the annuli nearest to the center of the galaxy. Radio mode feedback injects energy into the halo to prevent the cooling of hot gas, and is responsible for regulating star formation and the stellar mass of galaxies. Without this feedback, galaxies would accumulate substantial cold gas reserves, leading to excessively high star formation rates and resulting in galaxies with greater stellar masses than observed (see e.g. \citealt{Croton2006}).
\par
Additionally, {\sc Dark Sage} has three star formation channels: star formation from molecular gas; merger-induced starbursts; and Toomre instabilities. The first star formation channel is dependent on the local $H_2$ content defined as:

\begin{equation}
\Sigma_{\rm SFR}(r) = \epsilon_{\rm SF}\, \Sigma_{\rm H_2}(r)~, \label{eq:DSSFR_eq} 
\end{equation}
where $\epsilon_{\rm SF}$ is the star formation efficiency and $\Sigma_{\rm SFR}(r)$ is the molecular hydrogen surface density which is calculated within each annulus. In {\sc Dark Sage}, the cold gas disk includes hydrogen, helium, and some ionized gas. The second star formation channel acts upon a galaxy merger. A merger-driven starburst happens whenever gas from two galaxies collides within an annulus. The third star formation channel occurs when an annulus undergo starbursts to resolve unstable gas from secular processes or leftover unstable gas from merger-induced starbursts. After galaxies merge, gas that is left from starbursts is transferred to neighboring annuli. This would trigger a star formation episode to attempt resolve instabilities within the annulus.
\par
Star formation episodes are tied to supernova feedback. Supernovas can heat the gas, and the amount of energy injected depends on the local heating surface density of the gas:

\begin{equation}
\textstyle\sum\nolimits_{\rm reheated}(r) = \epsilon_{\rm disk} \ \frac{\textstyle\sum_{\rm 0,gas}}{\textstyle\sum_{\rm gas}(r)} \ \textstyle\sum\nolimits_{\rm SFR}(r)
\label{eq:SN_mode}
\end{equation}

where $\textstyle\sum_{\rm 0,gas}$ is the reference surface density, $\textstyle\sum_{\rm gas}$ is a gas surface density, $\textstyle\sum\nolimits_{\rm SFR}(r)$ is obtained from eq. \ref{eq:DSSFR_eq} and the $\epsilon_{\rm disk}$ parameter is the mass-loading factor. As a lower limit for resolution purposes, supernova feedback occurs if a star formation event within an annuli produces stars with mass greater than $100 \mathrm{M}_{\odot} h^{-1}$. 
\par
For a comprehensive overview of {\sc {\sc Dark Sage}}, we refer the reader to \citet{Stevens+2016}.
\section{Comparing simulations with the semi-analytic model}
\label{sec:critera}
In order to robustly compare these two simulation suites and the semi-analytic model {\sc Dark Sage}, we discuss some general points related to data selection and cutoffs:
\begin{enumerate}
    \item \textbf{Centrals}: we only consider central galaxies in all the data-sets.
    \item \textbf{Galaxy/halo mass cut}: for all datasets, we restrict our samples to $M_* \geq 10^9 M_\odot$. This ensures that galaxies in {\sc Astrid} and {\sc TNG300} are resolved with sufficient number of star particles ($\sim 2500$ for {\sc Astrid}, $\sim 1000$ for TNG300). For {\sc Dark Sage}, we impose an additional halo mass cut $M_h(z=0) \geq 10^{11.2} M_\odot$, which ensures that each halo at $z=0$ is resolved by at least $\sim 200$ particles. 
    \item \textbf{Black hole mass}: for all datasets at all redshifts, we use a black hole mass cutoff of $M_{\rm BH} \geq 10^6 M_\odot$. This cut is imposed so that our results on BH growth and its interplay with star formation are less sensitive to the exact seeding prescription adopted. 
    \item \textbf{Black hole occupation}: for the simulations, we only consider galaxies with at least one BH, and we report only the most massive BH hosted by each galaxy. For massive galaxies, the most massive BH is almost always located close to the subhalo center. However for lower mass galaxies in {\sc Astrid}, the most massive BH could be off-center. We define a BH to be a ``central BH" if its offset from the subhalo center of mass is $\leq 1$ kpc. In {\sc Dark Sage}, all galaxies are seeded with a BH fixed at the center, hence all BHs are central. 
    \item \textbf{Stellar mass and star formation rate}: all stellar masses are the total stellar mass of the subhalo for the simulations. Star formation rates are the integrated star formation rate of every gas particle of the subhalo in the simulations. For {\sc Dark Sage}, the stellar masses/star formation rates are the integrated quantities over all annuli.   
\end{enumerate}

\section{Results comparing simulations with the semi-analytic model}
\label{sec:mbh_mstar}
\subsection{Mean  $M_{\rm BH}-M_*$ relation}

We first examine the evolution of the $M_{\rm BH}-M_*$ relation across redshift. We calculate the median value of $M_{\rm BH}$ for the galaxies satisfying the criteria in section \ref{sec:critera} within each stellar mass bin. We use bins of width 0.25 dex between $10^9-10^{12} M_\odot$. We have verified that the results using the mean relation are very similar to the median relation. The results are also largely independent of the choice of bins (as long as the bin sizes are not too small, i.e. $\lesssim 0.1$ dex).
\par
\begin{figure*}
    \centering
    \includegraphics[width=\textwidth]{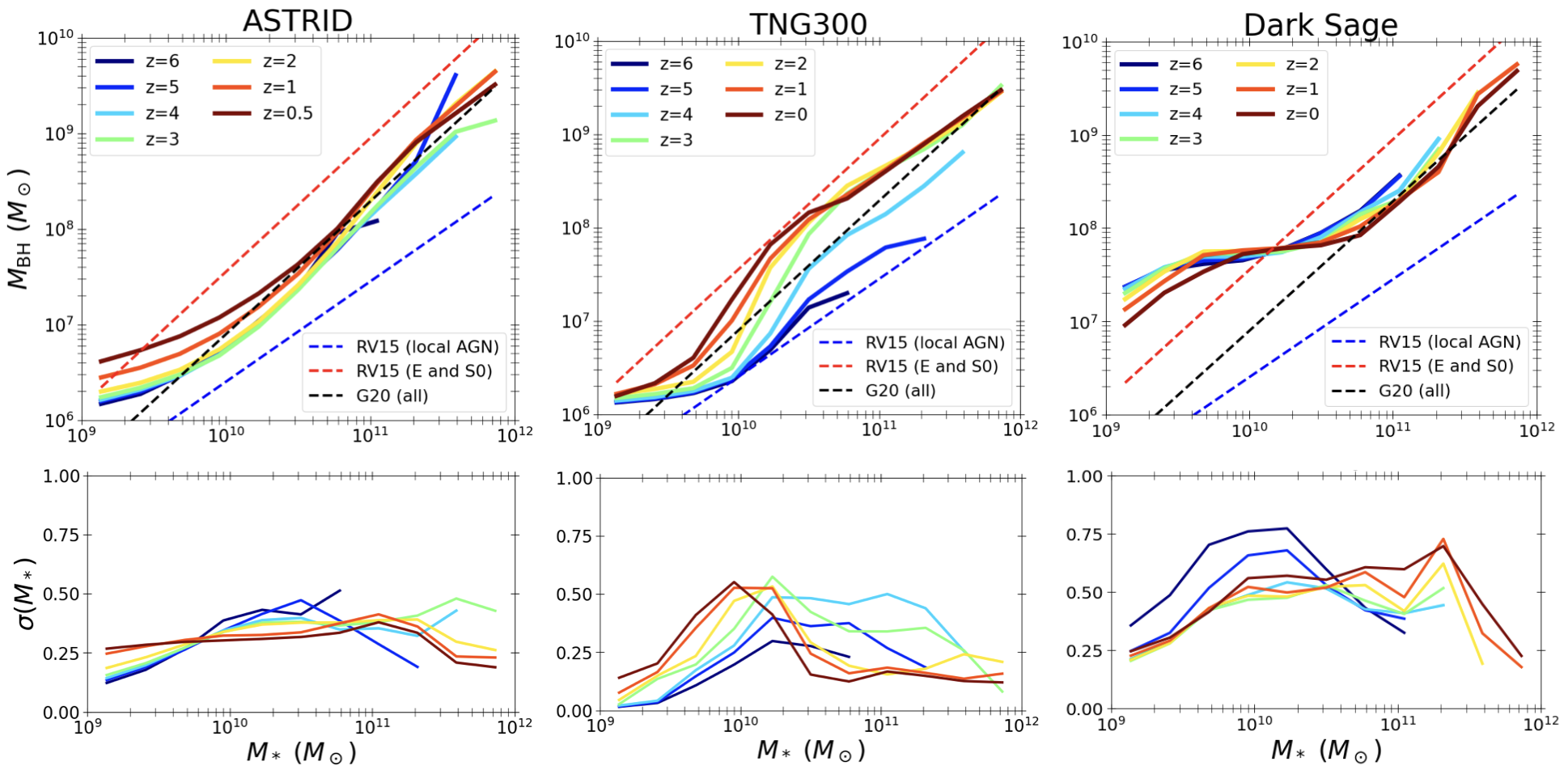}
    \caption{Top: median $\mbh-M_*$ relations in {\sc Astrid} (left), TNG300 (center), and {\sc Dark Sage} (right). We also show the empirical relations for the $\mbh-M_*$ relation for local AGN and local ellipticals and classical bulges from \citet{Reines2015}, and the average local relation for all types from \citet{Greene2020}. {\sc Astrid} and {\sc Dark Sage} show little to no evolution in the $\mbh-M_*$ relation from $z=6$ to the lowest redshift, while TNG300 shows a significant upward evolution over time. Bottom: scatter in $\mbh-M_*$, defined by equation \ref{eq:scatter}.}
    \label{fig:mean_mbh_mstar}
\end{figure*}

The top panel of Figure~\ref{fig:mean_mbh_mstar} shows the median $M_{\rm BH}-M_*$ relation across redshift for {\sc Astrid} (left), TNG300 (middle), and {\sc Dark Sage} (right).  We overplot the $M_{\rm BH}-M_*$ empirical relation for local ellipticals and local AGN from \citet{Reines2015}, and the average relation for all galaxies from \citet{Greene2020}. We note caution when comparing the simulation results to observations, as our definition of $M_*$ is the total stellar mass within the sub-halo. Meanwhile from the observations, \citet{Reines2015} and \citet{Greene2020} report the total stellar mass of the galaxy. This can offset the stellar masses in our analysis towards higher values; however this offset does not affect our conclusions regarding the \textit{direction} of evolution of the $\mbh-M_*$ relation.
\par
We note the following features in our $\mbh-M_*$ relations from simulations and {\sc Dark Sage}:

\begin{enumerate}
    \item At the $z=0$, the median $\mbh-M_*$ relations in all three datasets lie in-between the \citet{Reines2015} relations for active and inactive galaxies, and are in close agreement with the \citet{Greene2020} relation. We include all galaxies in our median $\mbh-M_*$ calculation and do not separate active and inactive galaxies. Several studies (e.g. \citealt{Reines2015, Sahu2019, Greene2020}) show that the normalization of the $\mbh-M_*$ scaling relation is higher for early-type galaxies compared to late-type galaxies. Therefore, it is expected that the median relation for the total set of galaxies lies between the relations for the active and inactive populations.
    \item {\sc Astrid} and {\sc Dark Sage} show little to no evolution in the $\mbh-M_*$ relation from $z=6$ to the lowest redshift bin. On the other hand, TNG300 presents a $\mbh-M_*$ relation that moves upwards over time. For $z \lesssim 2$, observations point towards either no evolution in the $\mbh-M_*$ relation \citep{Shields2003, Sun2015, Suh2020, Li2023} or a downwards evolution over time \citep{Merloni2010, Bennert2011, Ding2020}. At higher redshifts, the relation is largely unconstrained, with JWST results pointing to over-massive BHs compared to the local relation \citep{Maiolino2023, Harikane2023, Pacucci2023, Mezcua+2024} (but see \citealt{Li2024}).
    \item At low stellar masses ($M_* \lesssim 10^{10} M_\odot$), the $\mbh-M_*$ relation flattens at all redshifts for {\sc Astrid} and at high redshift for TNG300. This is attributed to the effect of supernova feedback that suppresses BH growth preferentially in low-mass halos \citep{Dubois2015, Angles2017, Habouzit2017, Habouzit2021}. Below a characteristic mass of $M_* \sim 10^{10} M_\odot$, SN winds are able to eject gas out of the galaxy entirely and with this evacuation of gas, BHs cannot grow efficiently. A similar flattening of the relation is seen in {\sc Dark Sage}, but this effect is largely numerical. At $z=0$, a population of galaxies below $M_* \sim 10^{10} M_\odot$ also lie below our halo mass cut ($M_h \sim 10^{11.2} M_\odot$). Therefore, the high number of low-mass galaxies missing may bring down the $\mbh-M_*$ relation below this characteristic mass. 
\end{enumerate}

\subsection{Evolution of the scatter in the $\mbh-M_*$ relation}

The bottom panels of Figure~\ref{fig:mean_mbh_mstar} show the evolution in the scatter in the $\mbh-M_*$ relation across redshift. We define scatter as:
\begin{equation}
\label{eq:scatter}
    \sigma(M_{\rm BH} | M_*) = \left( \log_{10} M_{\rm BH,84}-\log_{10}M_{\rm BH,16} \right)/2
\end{equation}
where $M_{\rm BH,16}$ and $M_{\rm BH,84}$ are the 16 and 84 $\mbh$ percentile values (in units of $M_\odot$) in a given $M_*$ bin.  \footnote{Our selection criteria includes $\mbh \gtrsim 10^6 M_\odot$, so the reported scatter (especially at the low mass end) is smaller than the total scatter in {\sc Astrid} and {\sc Dark Sage}, when including populations with $\mbh \lesssim 10^6 M_\odot$.} 
\par

Out of the three datasets, {\sc Dark Sage} has a significantly larger scatter than {\sc Astrid} and TNG300 at most values of redshift and $M_*$. 
\citet{PorrasValverde2023} show that {\sc Dark Sage} (as well as other SAMs) has a larger scatter in the stellar mass-halo mass relation than TNG300. {\sc Dark Sage} also implements multiple modes of BH growth (see section \ref{sec:datasets}). These factors could contribute towards a larger $\mbh-M_*$ scatter. The two cosmological simulations have similar levels of scatter at intermediate masses. However, {\sc Astrid} has a larger scatter than TNG300 at the low mass end at high redshift, which could be partly due to a distribution of seed masses instead of a fixed seed mass. We discuss this further in section \ref{sec:discussion}. 
\par
Directly comparing the scatter in the derived $\mbh-M_*$ relation between simulations/SAMs and observations is challenging. The total scatter in observations consists of both the intrinsic scatter and the uncertainties in mass estimates. Moreover, it is necessary to account for sample completeness and detection limits in order to obtain the ``true" scatter from an observational sample. Keeping these points in mind, the scatter (as defined by the 15-85 percentile values) in the \citet{Reines2015} sample between $M_* \sim 10^{9.5}-10^{11} M_\odot$ is $\sigma \sim 0.9-1.9$, which is higher than either {\sc Astrid} or {\sc TNG300}. 

\subsection{Interplay between BHAR and SFR}

The $\mbh-M_*$ scaling relation and its evolution depends on the growth rates of both the black holes and the stellar content of their host galaxies. The majority of the mass budget for $\mbh$ and $M_*$ comes from BH accretion and star formation respectively, with little contribution from mergers \citep{Hirschmann2010, Pacucci2020}. Therefore in this section, we study the interplay between the black hole accretion rate (BHAR) and the star formation rate (SFR) across redshift and stellar mass. 

\par

\begin{figure*}
    \centering
    \includegraphics[width=\textwidth]{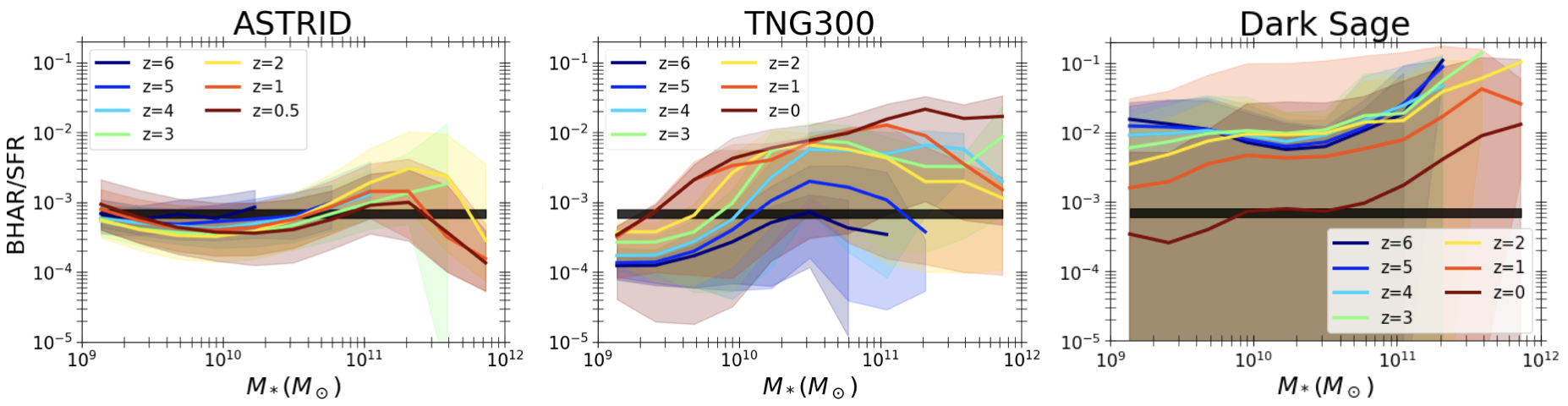}
    \caption{Median values of BHAR/SFR vs stellar mass in {\sc Astrid}, TNG300, and {\sc Dark Sage}. The shaded regions show the $1-\sigma$ intervals. We also show the $z\sim 1$ relation from \citet{Mullaney2012} in black. All the three models show different trends of BHAR/SFR with $M_*$ and across redshift.}
    \label{fig:bhar_sfr}
\end{figure*}

Figure \ref{fig:bhar_sfr} shows the median value and 16-84 percentile range of the BHAR/SFR ratio in each stellar mass bin in {\sc Astrid} (left), TNG300 (middle), and {\sc Dark Sage} (right). We also show the relation from \citet{Mullaney2012} at $z \sim 1$. From figure \ref{fig:bhar_sfr}, {\sc Astrid} has an average BHAR/SFR ratio that is mostly constant over redshift and stellar mass. This reflects the non-evolving and non-linear $\mbh-M_*$ relation in {\sc Astrid} from figure \ref{fig:mean_mbh_mstar}.
\par
In {\sc TNG300}, the BHAR/SFR ratio rises with both increasing stellar mass ($M_*$) and higher redshifts. This is in line with the upwards evolution of the $\mbh-M_*$ relation in TNG300. {\sc Dark Sage} shows a similar trend of increasing BHAR/SFR with $M_*$, but shows downward redshift evolution for $z\leq 2$. Interestingly, {\sc Dark Sage} shows BHAR/SFR ratios at $z>2$ that are significantly higher than the \citet{Mullaney2012} relation even though the $\mbh-M_*$ relation is still in agreement with local values for $M_* \gtrsim 10^{10} M_\odot$. However, the BHAR/SFR ratio is doen not evolve much for $z \gtrsim 2$, which is in line with the non-evolving $\mbh-M_*$ relation in {\sc Dark Sage}. In {\sc Dark Sage} at $z \gtrsim 2$, instabilities and merger-induced BH accretion contribute to most of the BH mass growth, therefore BHs are growing at a higher rate than their host galaxies (Porras-Valverde et al. submitted). By $z<2$, the feedback is so strong that it levels out the SFR with the BH accretion although BHs still grow through BH-BH mergers. 
\par
We find that the scatter in the ratio of BHAR/SFR is significantly higher in {\sc Dark Sage} compared to {\sc Astrid} and TNG300, with {\sc Astrid} having the least scatter. This higher BHAR/SFR scatter in {\sc Dark Sage} is turn reflected in the high scatter seen in the $\mbh-M_*$ relation. We discuss the implications of this in section \ref{sec:discussion}.

\section{Interplay between BH accretion and star formation rates}
\label{sec:bhar_sfr}

\subsection{Predicting the evolution of the $\mbh-M_*$ relation}

Given the strong correlation between the globally averaged SFR and BHAR, we now further study how the interplay of BHAR and SFR traces the $\mbh-M_*$ relation. We first present a simple toy model relating the relevant quantities. Assuming that a BH is growing perfectly in tandem with its host galaxy as a function of time, we have:
\begin{equation}
    \log \left( \mbh/M_\odot \right) = \alpha \log \left( M_*/M_\odot \right) + \beta
\end{equation}
We take $\alpha=1$, i.e. $\mbh$ is proportional to $M_*$ at all times. Differentiating with respect to time, we get:
\begin{equation}
    \frac{\dot{M}_{\rm BH}}{M_{\rm BH}} = \frac{\dot{M}_*}{M_*}
\end{equation}
If we ignore growth by mergers, we can set $\dot{M}_{\rm BH} = \rm{BHAR}$ and $\dot{M}_* = \rm{SFR}$. Therefore, if a BH is growing in exact proportion with its galaxy, we have:
\begin{equation}
    \mathcal{R} \equiv \rm{\frac{sBHAR}{sSFR}} = 1
\end{equation}
where $\rm{sBHAR}=\rm{BHAR/\mbh}$ and $\rm{sSFR}=\rm{SFR/M_*}$ are the specific BH accretion rate and specific star formation rate respectively. 
\par
If $\mathcal{R}>1$, then the BH is growing faster than its host galaxy, so the source is moving upwards in the $\mbh-M_*$ plane. This means that an under-massive BH (with respect to the mean relation) with $\mathcal{R}>1$ is moving towards the mean relation, whereas an over-massive BH is moving further away from the mean relation. Conversely, $R<1$ means that the galaxy's stellar mass is growing faster than its BH is assembling, so the source is moving rightwards in the $\mbh-M_*$ plane. 
\par
\begin{figure*}
    \centering
    \includegraphics[width=\textwidth]{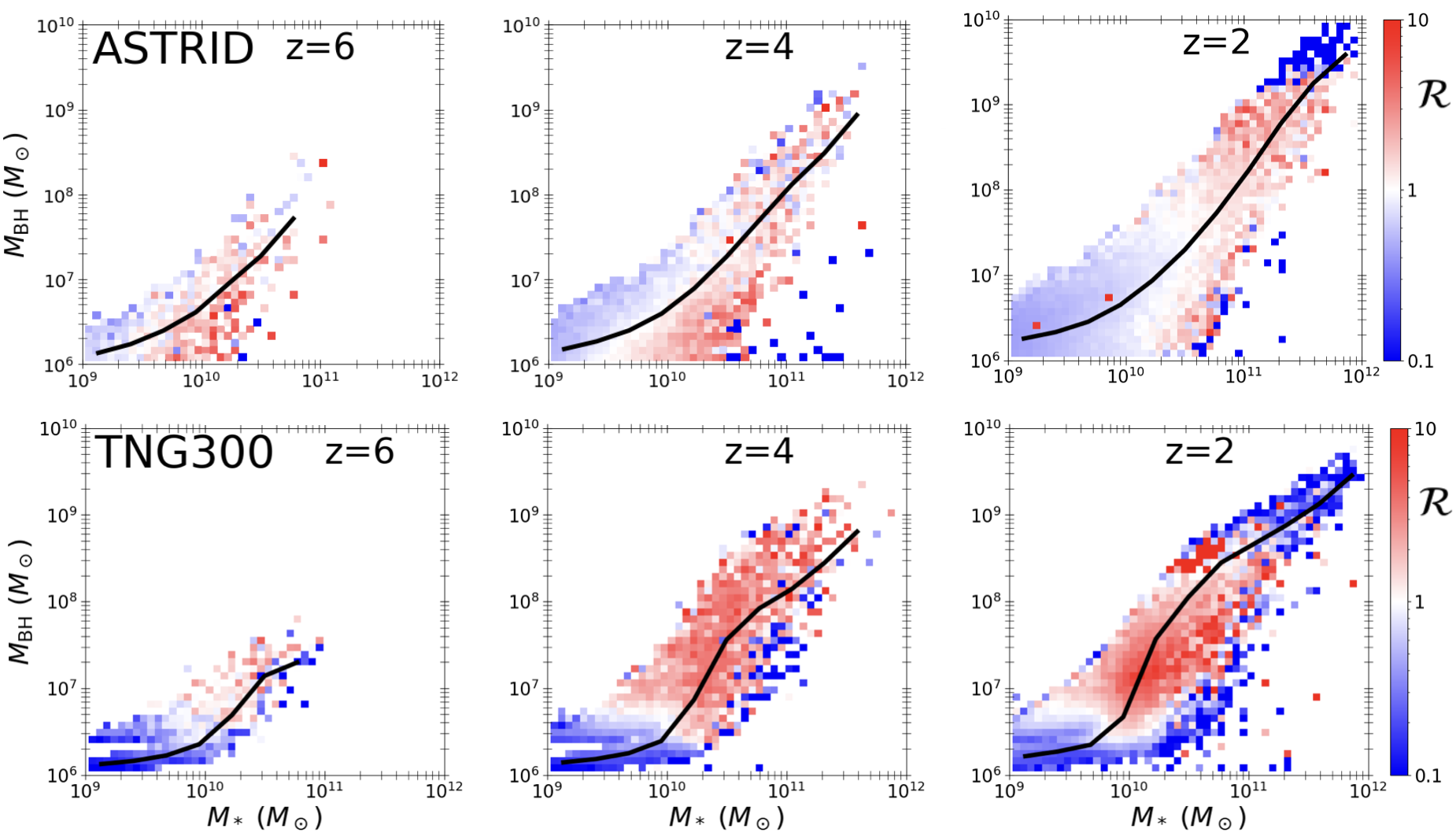}
    \caption{Top panels: $M_{\rm BH}-M_*$ color coded by the values of $\mathcal{R}=$sBHAR/sSFR for {\sc Astrid} at $z=6$ (left), $z=4$ (center), and $z=2$ (right). The median $\mbh-M_*$ relation is shown in black. Roughly, sources with $\mathcal{R} \geq 1$ tend to lie below the median relation, whereas sources with $\mathcal{R} \leq 1$ tend to lie above the mean relation. Bottom panels: same plots for {\sc TNG300}. Unlike {\sc Astrid}, a large population of galaxies have $\mathcal{R} >1$, regardless of whether the BHs are under-massive or over-massive.}
    \label{fig:R_all}
\end{figure*}
In the top panels of Figure~\ref{fig:R_all}, we show distribution of $\R$ values in {\sc Astrid} in each $\mbh-M_*$ bin at $z=6$ (left), $z=4$ (center), and $z=2$ (right). The median $M_{\rm bh}-M_*$ relation at each redshift is shown in black. In {\sc Astrid}, BHs that are over-massive with respect to the mean relation tend to have $\mathcal{R}<1$, whereas BHs that are under-massive tend to have $\mathcal{R}>1$. This indicates that on average, under-massive BHs are growing faster than their host galaxies, whereas over-massive black holes are growing slower than their host galaxies. Therefore systems are on average moving towards the mean relation, which is in line with the non-evolving nature of the $M_{\rm BH}-M_*$ relation. We demonstrate this co-evolution by tracing the growth histories of BHs in appendix \ref{sec:growth_history}. Note that despite these average trends, there is scatter in the $\mathcal{R}$ values among individual sources. 
\par
When we repeat this analysis for TNG300 as shown in the bottom panels of figure \ref{fig:R_all}, $\mathcal{R}>1$ sources dominate the entire population at $z=6,4,$ and $2$ (except the massive end at $z=2$). Therefore for galaxies in the $M_* \lesssim3 \times 10^{11} M_\odot$ range, their BHs are on average growing faster than their host galaxies. This is reflected by the upwards evolution of the median $\mbh-M_*$ relation. We note that at the most massive end ($M_* \gtrsim 5 \times 10^{11} M_\odot$), both the sBHAR and sSFR are low due to quenching from kinetic feedback. Therefore most BH/stellar mass growth for these galaxies occurs primarily through mergers, which is not reflected in the $\mathcal{R}$ value. 

\subsection{The specific star formation rate -- specific BH accretion rate plane}
\label{sec:quadrants}

\begin{figure}
    \centering
    \includegraphics[width=\columnwidth]{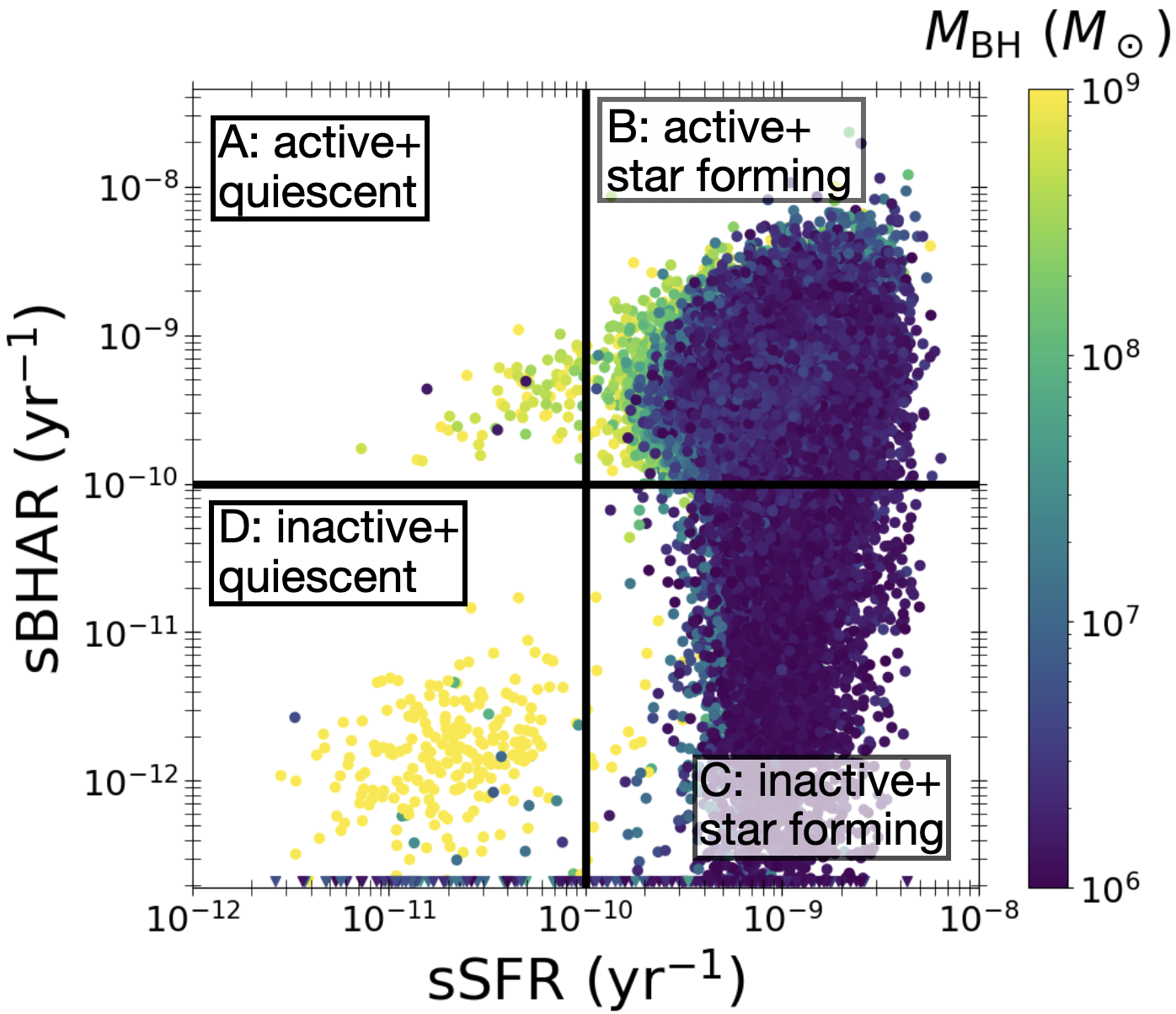}
    \caption{{\sc Astrid} galaxies in the sSFR-sBHAR plane at $z=2$, color coded by the value of $\mbh$. We divide the sSFR-sBHAR into four quadrants as shown. All four quadrants are occupied, demonstrating the interplay between BH accretion and star formation in the simulation. }
    \label{fig:z2_all}
\end{figure}

We now turn to study the physical mechanisms that govern the interplay between SFR and BHAR. We focus on {\sc Astrid} since it has the highest resolution, and its subgrid dynamical friction model allows for wandering BHs which leads to interesting features, as we shall shortly see. Figure \ref{fig:z2_all} shows the distribution of all galaxies in {\sc Astrid} at $z=2$ in the sSFR-sBHAR plane. The points are color coded by the value of $\mbh$. Roughly, we can divide the population into four quadrants. In figure \ref{fig:z2_misc} in appendix \ref{sec:quad_misc}, we show the same data color coded by different quantities to further illuminate the differences between the quadrants. We examine the properties of the galaxies and their BHs in each quadrant below, and describe how such demarcation furthers our understanding of the nature of quenching. 
\begin{itemize}
    \item \textbf{Quadrant A: active+quiescent}  \\
    \\
    \boxed{{\rm sSFR} \leq 10^{-10} \ {\rm yr}^{-1}$ and ${\rm sBHAR} \geq 10^{-10} \ {\rm yr}^{-1}}  \\ 
    \\
    This quadrant is relatively less populated and contains galaxies which are not highly star forming but host BHs that are actively accreting. In the $\mbh-M_*$ plane, these galaxies have BH masses that are overmassive compared to the mean relation. These galaxies are relatively compact (figure \ref{fig:z2_misc}) and are characterized by ``outside-in'' quenching, as we will show in \S\ref{sec:quench}. 
    
    \item \textbf{Quadrant B: active+star forming}  \\
    \\
    \boxed{{\rm sSFR} \geq 10^{-10} \ {\rm yr}^{-1}$ and ${\rm sBHAR} \geq 10^{-10} \ {\rm yr}^{-1}}  \\ 
    \\
    This quadrant predominantly hosts gas-rich, low-to-intermediate mass galaxies with $\mbh \leq 10^{8.5} M_\odot$. These BHs are not effective at quenching their hosts since they have not yet turned on their kinetic feedback mode. Therefore, their galaxies have high values of both sSFR and sBHAR. 
    
    \item \textbf{Quadrant C: inactive+star forming}  \\
    \\
    \boxed{{\rm sSFR} \geq 10^{-10} \ {\rm yr}^{-1}$ and ${\rm sBHAR} \leq 10^{-10} \ {\rm yr}^{-1}}  \\ 
    \\
    These galaxies are characterised by high star formation rates but low black hole accretion rates. This population is dominated by off-center ``wandering'' BHs, as we show in figure \ref{fig:z2_misc} (appendix \ref{sec:quad_misc}). These wandering BHs do not accrete efficiently \citep{Bellovary2019, Ricarte2021} and hence do not grow significantly above the seed mass. On the other hand, the star formation rate is unaffected by the presence of a wandering BH. However there are also 119 sources in this quadrant that host central (non-wandering) BHs, which we examine further in section \ref{sec:quench}.
    \par
    The number of galaxies in quadrant C depends crucially on the modelling of BH dynamics and dynamical friction. In {\sc TNG300}, where BHs are seeded at the halo center and instantaneously re-positioned at the center of the potential well, this population is absent for $z \geq 2$.

    \item \textbf{Quadrant D: inactive+quiescent}  \\
    \\
   \boxed{{\rm sSFR} \leq 10^{-10} \ {\rm yr}^{-1}$ and ${\rm sBHAR} \leq 10^{-10} \ {\rm yr}^{-1}}  \\
    \\
    These galaxies primarily host massive, gas-poor galaxies with $\mbh \geq 10^{8.5} M_\odot$. This value of $\mbh = 10^{8.5} M_\odot$ is the threshold where kinetic feedback is turned on in {\sc Astrid} \citep{Ni2024}. In cosmological simulations, kinetic feedback is the primary feedback mechanism responsible for ejecting gas and preventing star formation and BH accretion in massive galaxies \citep{Weinberger2017, Habouzit2019, Terrazas2020}. Therefore this quadrant corresponds to massive, gas-poor systems. 
    
\end{itemize}

\subsection{Different quenching pathways}
\label{sec:quench}

Quadrants A and C (active+quiescent and inactive+star forming, respectively) with central BHs (i.e. offset $\leq 1$ kpc) are both characterized by different types of quenching, either star formation (in quadrant A) or BH accretion (in quadrant C). The bottom middle panel of Figure \ref{fig:z2_misc} shows all the central BHs in the sSFR-sBHAR plane color coded by $\mbh$. The bottom right panel shows the same data color coded by stellar half-mass radius of the galaxy, $r_{1/2}$. From these plots, we see that quadrant A galaxies host massive BHs with $\mbh \gtrsim 10^{8.5} M_\odot$ and are compact ($r_{1/2} \sim 1$ kpc). On the other hand, quadrant C galaxies host BHs with $\mbh \lesssim 10^7 M_\odot$ and are more diffuse ($r_{1/2} \sim 2.5$ kpc).  
\par
\begin{figure*}
    \centering
    \includegraphics[width=0.49\textwidth]{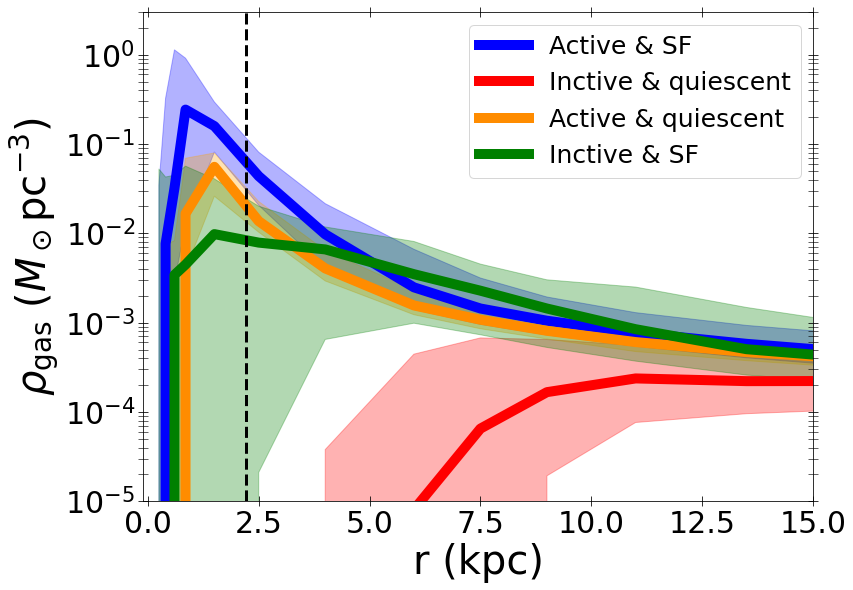}
    \includegraphics[width=0.49\textwidth]{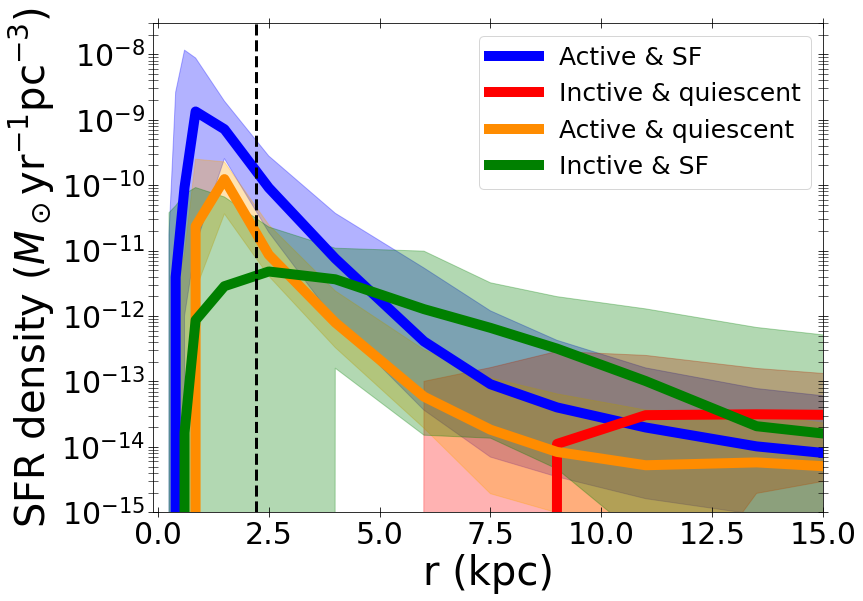}
    \caption{Total gas density profiles (left) and SFR density profiles (right) for {\sc Astrid} galaxies central BHs at $z=2$. The active+star forming galaxies have large gas reservoir and high SFR at all radii, whereas inactive+quiescent galaxies have little to no gas and star formation in the inner regions. The active+quiescent and inactive+star forming galaxies  show intermediate properties (see discussion in text).} 
    \label{fig:gas}
\end{figure*}

Both BH accretion and star formation depends on the presence of gas. In figure \ref{fig:gas}'s left panel, we plot the median gas density profiles and their 1-$\sigma$ range for random samples of galaxies from all four quadrants. While there is considerable scatter amongst individual galaxies, $\rho_{\rm gas}$ in the outer regions ($r \gtrsim 3$ kpc) is similar between the two populations. However within the inner $\sim 3$ kpc, $\rho_{\rm gas}$ is roughly an order of magnitude higher for the active+quiescent galaxies compared to the inactive+star-forming galaxies. This high $\rho_{\rm gas}$ in the inner region allows for high accretion rates onto the BH.
\par
The right panel of Figure \ref{fig:gas} shows the median spherically averaged SFR density profiles for the same of galaxies. The higher $\rho_{\rm gas}$ in the inner $\sim 3$ kpc is reflected as a higher SFR density within this region. Outside of $\sim 3$ kpc, the average SFR density of the active+quiescent galaxies is over an order of magnitude lower than the inactive+star-forming galaxies. Even though the total gas density is the same in the outer regions, the star forming gas (i.e. cold gas) is lower for the active+quiescent galaxies, resulting in a lower sSFR. 
\par

\par

\subsection{Evolution of galaxies in the sSFR-sBHAR plane}

\begin{figure}
    \centering    \includegraphics[width=0.5\textwidth]{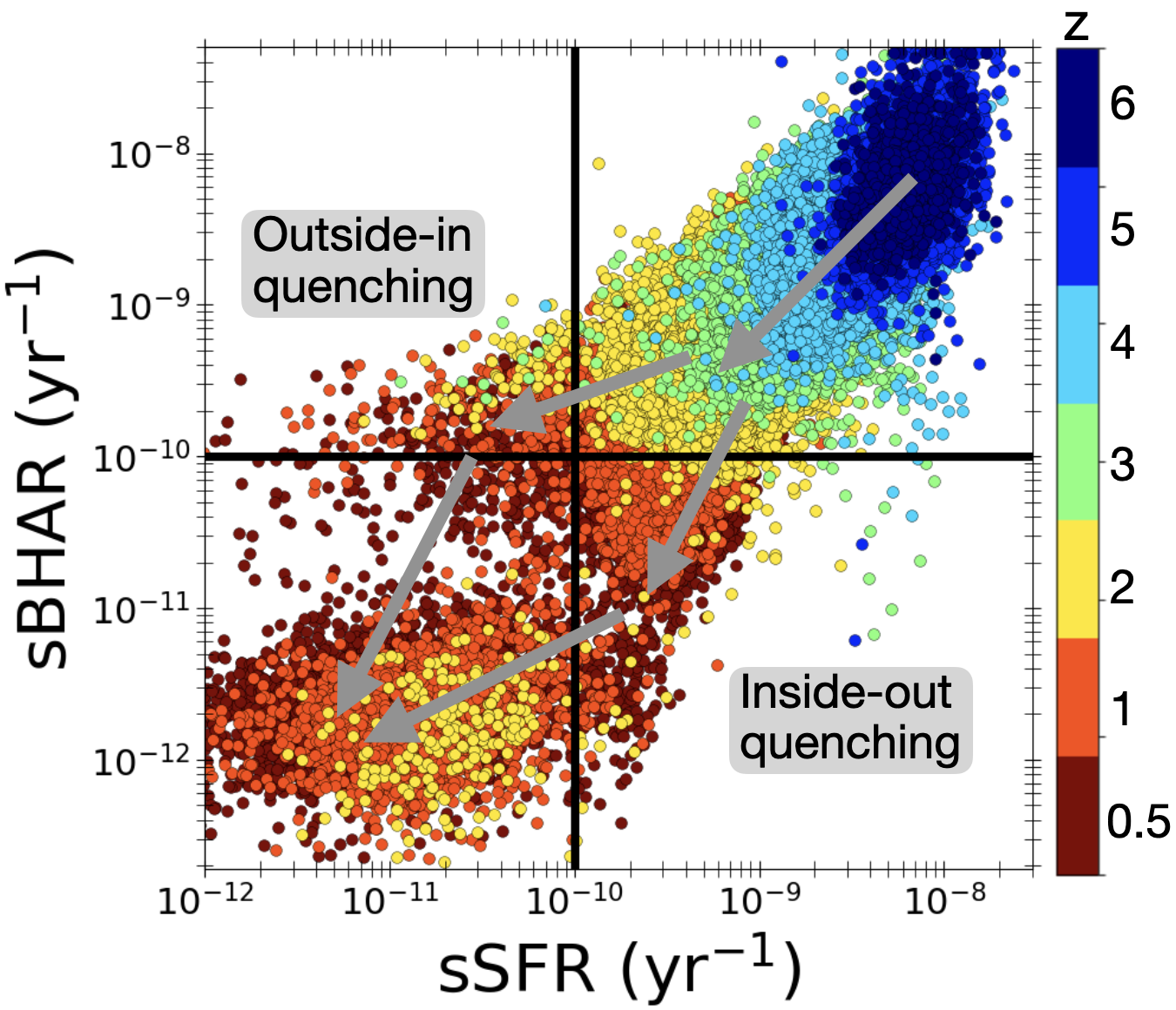}
    \caption{Tracks for {\sc Astrid} galaxies in the sSFR-sBHAR plane. Galaxies start at the top right corner at high redshift and move towards the bottom left as they grow and deplete their gas reservoir. Kinetic feedback enables the jump from quadrant B to D. Depending on the nature of quenching, galaxies can spend time in the ``intermediate" quadrants A or C. }
    \label{fig:direction}
\end{figure}

We now study how galaxies move in the sSFR-sBHAR plane over time. We select a population of galaxies at $z=6$ and trace their evolution down to $z=0.5$ (the lowest redshift in {\sc Astrid}). The redshift evolution of these galaxies in the sSFR-sBHAR plane is shown in figure \ref{fig:direction}, where each color represents a different redshift. At high redshift, when galaxies are gas rich and the stellar/BH masses are low, galaxies start out at the top right corner of quadrant B. This is qualitatively consistent with observational studies of high-redshift galaxies \citep{Aird2012, Lusso2012} and in other simulations \citep{Thomas2019, Habouzit2022} which show that high-redshift galaxies show higher values of sSFR and sBHAR. Over time, sources start moving towards the lower left as $\mbh$ and $M_*$ increase. Once the gas reservoir of the galaxy is depleted or ejected, both sSFR and sBHAR decrease by $\sim 2$ orders of magnitude. Since star formation and BH accretion occur on different spatial and temporal scales, it is unlikely that the sSFR and sBHAR values decrease at exactly the same rate. Therefore, the galaxy moves through quadrant A (outside-in quenching) or quadrant C (inside-out quenching). 

\section{Discussion}
\label{sec:discussion}

\subsection{Connecting black hole growth to star formation}

Since the same cold gas that goes into star formation also fuels the SMBH, one expects that the SFR should trace the BHAR. Indeed, both observations \citep{Aird2012, Mullaney2012, Chen2013} and simulations \citep{Ricarte2019, Thomas2019, Ni2022} show that the globally average SFR and BHAR are well correlated. However for individual galaxies, the observational evidence is less clear. The SFR of a galaxy evolves on the timescale of $\sim$100 Myr. But the BHAR might change on a significantly shorter timescales as reflected by AGN variability which operates on timescales of hours - Myrs \citep{Ulrich1997}. Therefore for an individual galaxy, the SFR and BHAR can appear to have little to no correlation \citep{Hickox2014, Angles2015}. 
\par
In addition, the physical scales relevant for the two quantities differ by multiple orders of magnitude, which can weaken the observed correlation between the global SFR and the BHAR. For example, \citet{Hopkins2010} find that the BHAR is more strongly correlated with the star formation rate at nuclear scales than global scales. Most cosmological simulations and SAMs only resolve BH accretion at the Bondi radius, which is $\sim 6$ orders of magnitude larger than the radius of the event horizon. The flow between the Bondi radius and sub-pc scales is subject to cold chaotic accretion due to the onset of thermal instabilities \citep{Gaspari2013}. Moreover, at sub-pc scales, the effects of magnetic fields and general relativity become important \citep{Cho2023,Cho2024}. For example, \citet{Cho2024} find that the accretion rate onto the BH is only $\sim 1 \%$ of the Bondi rate in their 3D GRMHD simulations. 
\par
The ratio between the BHAR and SFR of a galaxy gives the relative efficiency of fueling gas into the BH vs into stars. This ``AGN main sequence" \citep{Mullaney2012} traces out the nature and evolution of the $\mbh-M_*$ relation over time. However, the relationship between BHAR, SFR, and $M_*$ may be more complicated than a simple linear relation \citep{Delvecchio2015, Yang2017}. Figure \ref{fig:bhar_sfr} shows that the BHAR/SFR evolution and dependence on $M_*$ is different for our three datasets. This is reflected in the differences in the predicted $\mbh-M_*$ relation and its scatter. Moving forward, focusing on the BHAR-SFR interplay may help bridge the gap between observations and predictions from simulations/SAMs.

\subsection{Nature of quenching}

JWST has revealed a large population of massive quenched galaxies at high redshift \citep{Labbe2023,Carnall2023, Nanayakkara2024}. It is believed that AGN feedback is the primary quenching mechanism for massive galaxies and at high redshift \citep{Pacucci2024}. In a recent follow-up study of JWST sources at cosmic noon, \citet{Park+2023} find that rapid quenching appears to be driven by AGN feedback. In simulations, kinetic feedback is significantly stronger at quenching galaxies compared to thermal feedback \citep{Terrazas2020, Weller2024}. Both {\sc TNG300} and {\sc Astrid} only include kinetic feedback at high BH masses and/or low Eddington ratios, and only for $z < 2.4$ for {\sc Astrid}. This results in quenching only when the BH reaches a threshold mass of $\sim 10^8 M_\odot$, which typically happens at low redshift. Indeed, figure \ref{fig:direction} shows that there are very few quenched galaxies at $z \geq 4$. Similar results are seen in {\sc TNG300}.

\par
Quenched galaxies in simulations often host BHs that are over-massive compared to the mean relation \citep{Scharre2024, Weller2024, Ni2024}. Self-regulation of the host galaxy's gas supply via feedback limits the growth of these over-massive BHs \citep{Natarajan2009, Pacucci2017}. As such, over/under-massive BHs at high redshift need not remain over/under-massive at lower redshift \citep{DiMatteo2008, Volonteri2009}, and the host galaxies can undergo reactivation of star formation at a later time. In future work, we plan to focus on the origin of over-massive BH galaxies and their subsequent evolution.
\par
In addition, different quenching mechanisms leave different observational signatures on their host galaxies (e.g. \citealt{Woo2019}). Some studies point towards ``inside out'' quenching being the dominant quenching mechanism due to AGN feedback \citep{Lin2019, Nelson2021}. This is characterized by star formation first shutting off in the central regions, resulting in older centers compared to outskirts. However, the opposite has also been observed (e.g. \citealt{Barro2016, Suess2021}), i.e. galaxies with younger centers compared to outskirts.
\par
\citet{Ni2024} extensively studied the population of quenched galaxies in {\sc Astrid} and found a correlation between the quenching pathway and galaxy morphologies. They find that massive compact galaxies often undergo a compaction-like quenching mechanism, which results in rapid quenching and younger central regions. This analogous to the ``outside-in" quenching pathway described in section \ref{sec:quench}. \citet{Ni2024} find that on the other hand, galaxies with more diffuse morphologies have longer quenching timescales and older central regions, which is the ``inside-out" quenching pathway. These quenching pathways are reflected in the different tracks in the sSFR-sBHAR plane (figure \ref{fig:direction}).

\subsection{Seeding vs dynamics vs feedback}
Section \ref{sec:bhar_sfr} shows that the tracks of BH growth and co-evolution with its host galaxy depends on the implementation of seeding, dynamics, and feedback. Disentangling the effects of these processes is a major goal of current work on BH-galaxy co-evolution. For example, {\sc Astrid} seeds BHs by converting the densest gas particle of a halo into a BH particle. While this often corresponds to a particle close to the center of mass, this need not always be the case. Zoom-in cosmological simulations \citep{Bellovary2021} find that a small fraction of BHs are formed off-center. Off-center BHs have to sink to the center of the potential well before they start accreting efficiently. This sinking timescale depends on the resolution and/or modelling of dynamical friction. 
\par
Feedback includes the contribution from supernovae and AGN, which itself is usually divided into thermal and kinetic modes. Differences in the implementation of any of these mechanisms can result in different co-evolution tracks \citep{Habouzit2021,Habouzit2022}. Controlled experiments to isolate the effects of the various processes at play will be the subject of our future work. 
\section{Conclusions}
\label{sec:conclusions}
We have analyzed the black hole-galaxy co-evolution in two cosmological simulations, {\sc Astrid} and TNG300, and the semi-analytic model, {\sc Dark Sage}. These models use different sub-grid recipes for BH seeding, dynamics, and feedback, hence showing different evolution tracks. Our key findings are as follows:
\begin{enumerate}
    \item Simulations currently disagree regarding the evolution of the $\mbh-M_*$ scaling relation. {\sc Astrid} and {\sc Dark Sage} show $\mbh-M_*$ relations that are mostly unevolving over time, whereas TNG300 shows a significant upward evolution in the relation (figure \ref{fig:mean_mbh_mstar}).
    \item The interplay between the BHAR and SFR guides the evolution of the $\mbh-M_*$ relation. The evolution of the mean BHAR/SFR ratio and its scatter is reflected in the $\mbh-M_*$ evolution. 
    \item The parameter $\mathcal{R}=$sBHAR/sSFR governs the direction in which a source moved in the $\mbh-M_*$ plane (towards or away the mean relation). By examining the mean $\mathcal{R}$ value across different galaxy populations, we can infer the direction in which the mean $\mbh-M_*$ relation moves. Due to different sub-grid recipes, different models differ in their distribution of $\mathcal{R}$ (figure \ref{fig:R_all}). 
    \item The distribution of sources in the sSFR-sBHAR plane can illuminate the some of the factors that drive the BH-galaxy co-evolution (figures \ref{fig:z2_all}, \ref{fig:direction}, \ref{fig:z2_misc}). 
    
\end{enumerate}
Overall, in this study we used three different simulation and SAM datasets to improve our understanding of the nature of AGN feedback and how it affects BH-galaxy co-evolution. In the future, we plan to examine the growth histories of individual galaxies of interest in {\sc Dark Sage}. In particular we will focus on the origin of over-massive BHs and quenched galaxies at high redshift. We also intend to examine the distribution of sources in the sSFR-sBHAR plane and their $\mathcal{R}$ values from surveys such as COSMOS. 
\begin{acknowledgments}
P.N. acknowledges support from the Gordon and Betty Moore Foundation and the John Templeton Foundation that fund the Black Hole Initiative (BHI) at Harvard University where she serves as one of the PIs. A.J.P.V gratefully acknowledges support of a post-doctoral fellowship from the Heising-Simons foundation. C.J.B. is supported by an NSF Astronomy and Astrophysics Postdoctoral Fellowship under award AST-2303803. This material is based upon work supported by the National Science Foundation under Award No. 2303803. This research award is partially funded by a generous gift of Charles Simonyi to the NSF Division of Astronomical Sciences. The award is made in recognition of significant contributions to Rubin Observatory’s Legacy Survey of Space and Time. 

\end{acknowledgments}

\software{}
\bibliography{sample631}{}
\bibliographystyle{aasjournal}

\appendix
\section{Growth histories of BHs}
\label{sec:growth_history}
\begin{figure*}
    \centering
    \includegraphics[width=\textwidth]{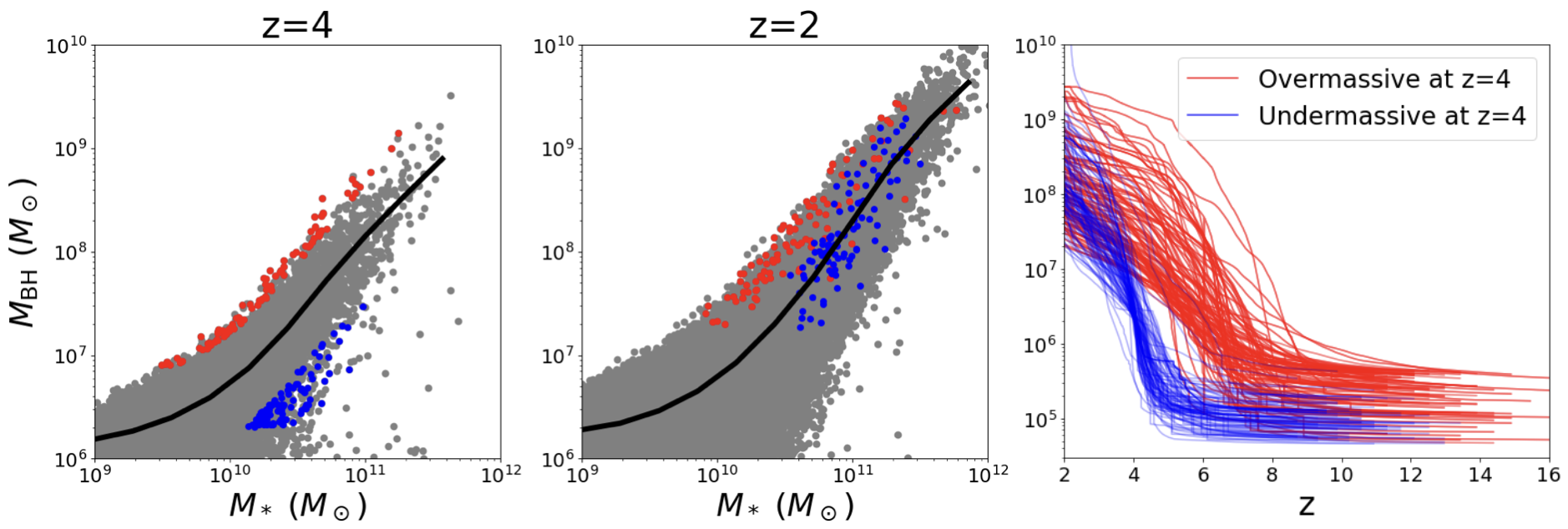}
    \caption{Tracing the growth histories of a sample of BHs in {\sc Astrid}. Left: the sample of under-massive (blue) and over-massive (red) BHs at z=4 in the $\mbh-M_*$ plane. Middle: the same BHs at $z=2$. Right: the growth histories of the sample of BHs from seeding to $z=2$. The under-massive BHs at $z=4$ have undergone a rapid burst of accretion, resulting in them lying along the median relation by $z=2$. On the other hand, the over-massive BHs at $z=4$ have significantly slower growth. }
    \label{fig:under_over}
\end{figure*}

Here, we examine the growth histories of samples of BHs in {\sc Astrid} to demonstrate co-evolution towards the mean relation. In figure \ref{fig:under_over}, we select a sample of 50 significantly ($\sim$ 0.5 dex) over-massive and under-massive BHs each in {\sc Astrid} at z=4. These sources are shown as red and blue points respectively in the $\mbh-M_*$ plane in the left panel of figure \ref{fig:under_over}. The over-massive sources all have $\mathcal{R} <1$ and the under-massive sources all have $\mathcal{R}>1$. 
\par
We then track their growth history from seeding to $z=2$. The middle panel shows the same sources at $z=2$, and the right panel shows the growth histories of all the BHs from seeding to $z=2$. BHs that were under-massive at $z=4$ have undergone a rapid growth phase, bringing them close to or even above the mean relation at $z=2$. On the other hand, the over-massive BHs at $z=4$ have a significantly slower growth between $z=4$ and $z=2$, bringing them closer to the mean relation (although most are still over-massive). 
\section{Quadrant properties}
\begin{figure*}
    \centering
    \includegraphics[width=\textwidth]{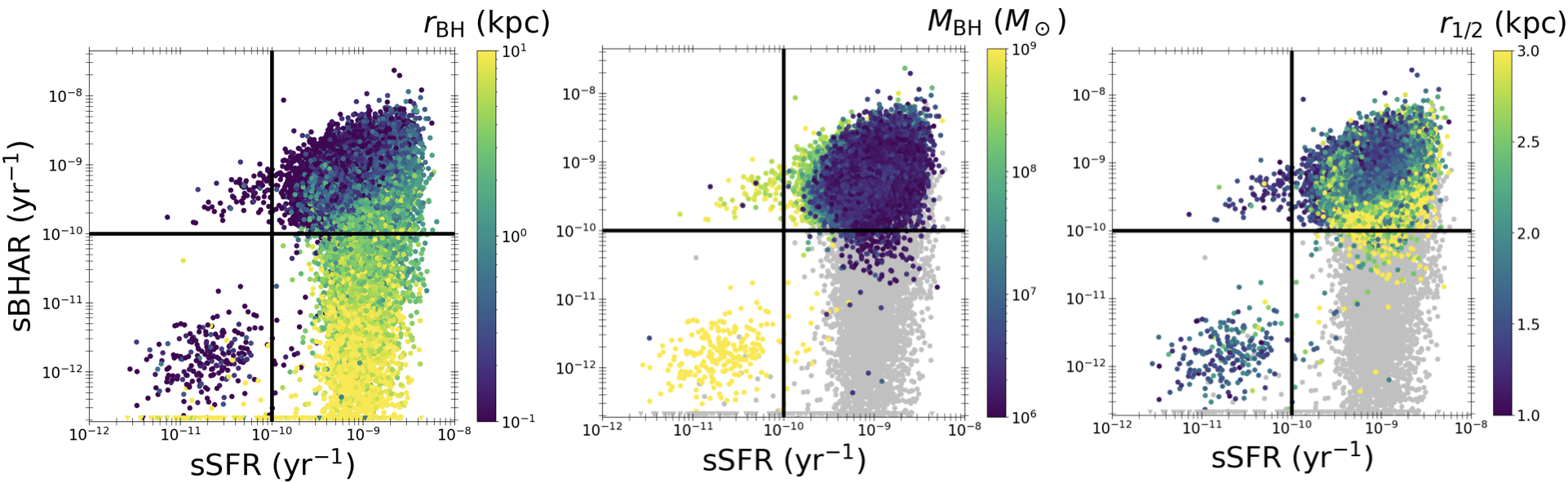}
    \caption{Left: same galaxies from figure \ref{fig:z2_all} in the sSFR-sBHAR plane color coded by $r_{\rm BH}$, the distance of the BH to the galaxy center. It is evident that the majority of the galaxies in the bottom right quadrant are off-center ($r_{\rm BH} \geq 1$ kpc). In the following two panels, the BHs with $r_{\rm BH} > 1$ kpc are grayed out. Center panel: central BHs ($r_{\rm BH} \leq 1$ kpc) color coded by $\mbh$. Right panel: central BHs color coded by the galaxy half-mass radius $r_{1/2}$. See section \ref{sec:quadrants} for discussion.}
    \label{fig:z2_misc}
\end{figure*}

In figure \ref{fig:z2_misc}, we show the same sources from figure \ref{fig:z2_all} in the sSFR-sBHAR plane, but color coded by different quantities in order to highlight the different properties of the quadrants. 
\par
Figure \ref{fig:z2_misc}'s left panel shows the galaxies color coded by $r_{\rm BH}$, the distance of the BH from the center of mass of the galaxy. The sources in the bottom inactive+star forming (right quadrant) mostly have high $r_{\rm BH}$. This prevents them from accreting efficiently, even if the galaxy is gas-rich and has a high sSFR. 
\par
In the center and right panels, we show the sources color coded by their black hole mass $M_{\rm BH}$ and half-mass radius $r_{1/2}$ respectively. We gray out the sources with $r_{\rm BH}>1$ kpc so we can focus on the central BHs. The average value of $M_{\rm BH}$ is highest in quadrant inactive+quiescent and lowest for quadrant active+star forming. In addition, the average $M_{\rm BH}$ is higher and $r_{1/2}$ is lower in quadrant active+quiescent compared to quadrant inactive+star forming. 
\label{sec:quad_misc}

\end{document}